\documentstyle[11pt,newpasp,epsf]{article}
\begin{document}

\title{Analysis of metallicity
effects  in SPH cosmological simulations}
\author{Tissera, ${\rm P.^1}$, Mosconi, ${\rm M.^2}$, Lambas, D. ${\rm G.^2}$}
\affil{1. Instituto de Astronom\'{\i}a y F\'{\i}sica del Espacio, Argentina}
\affil{2. Grupo de Investigaciones en Astronom\'{\i}a Te\'orica y
Experimental, Universidad Nacional de C\'ordoba, Argentina}

\begin{abstract}
We report preliminar results on a chemical model implemented
in a hydrodynamical cosmological code.
We compute the  metallicity of the gaseous component in 
galactic halos at different
redshifts. The results compare reasonably well to observations.
\end{abstract}

In recent years, there has been a dramatic 
improvement of our knowledge of the nearby
and high-z Universe. The cosmic star formation history
first depicted by Madau et al. (1998), together with observations of the
 ${\rm Ly_{\alpha}}$ forest, damped ${\rm Ly_{\alpha}}$ (DLA) systems and 
${\rm Ly_{\alpha}}$ break galaxies have brougth about information on
the chemical properties of the Universe at
different stages of evolution. 
Nevertheless, because the chemical properties of
galactic objects are directly linked
to their star formation (SF) histories which, on
their turn, are affected by various physical
processes (such as interactions, mergers, 
disk instabilities, etc), the information
they hold is of great relevance to deepen our understanding of
galaxy formation. Previous studies used  analytical chemical models
and were
 avocated to the Galaxy (Chiappini, Matteucci \& Graton (1997) and references there in).
There have been also some 
attempts to include chemical enrichment models
in numerical codes (Steinmetz \& Muller 1994; Raiteri, Villata \& Navarro 1996). 
In this work, we report first results of an 
implementation of a chemical enrichment model
in a hydrodynamical cosmological code based on
the Smooth Particle Hydrodynamics technique (Tissera, Lambas \& Abadi 1997).
These fully consistent cosmological simulations 
take into account the gravitational and hydrodynamical evolution
of the structure in a cosmological context.
Cold and dense gas particles are transformed into stars according to the
Schmidt law. We include the treatment of the metal production of
SN type I and II explosions.
We use the SPH technique to distribute
the metals within the neighboring sphere of
the particle where star formation has occurred. 
This approach is not a closed box model and
particles can be enriched by more than
one neighbour particle at each time-step as the move
according to the gravitation and hydrodynamics forces.

We perform tests of the effects of numerical resolution which show that
after the SF process starts the distributions of metals are very similar 
for objects resolved with 500, 1000 and 2000 gas  particles. The major impact is on the dispersion 
of metallicity:
the lower the number of gas particles, the larger the dispersion.
We run cosmological simulations consistent with
 a CDM model with $\Omega=1, \sigma_8=0.6$ in agreement
with observed cluster abundances using $N=64^3$ total particles ($10\%$ assumed
to be baryons).
Galactic halos are identified at their virial radius 
at different redshifts and their astrophysical and chemical properties 
analyzed as function of $z$. 
In particular, in Figure 1, we show the metal content of the gaseous component of halos
with total masses larger than $3.9 \times 10^{11} {\rm M_{\odot}}$,
at the virial radius, identified at different redshifts. 
\begin{figure}
\plotfiddle{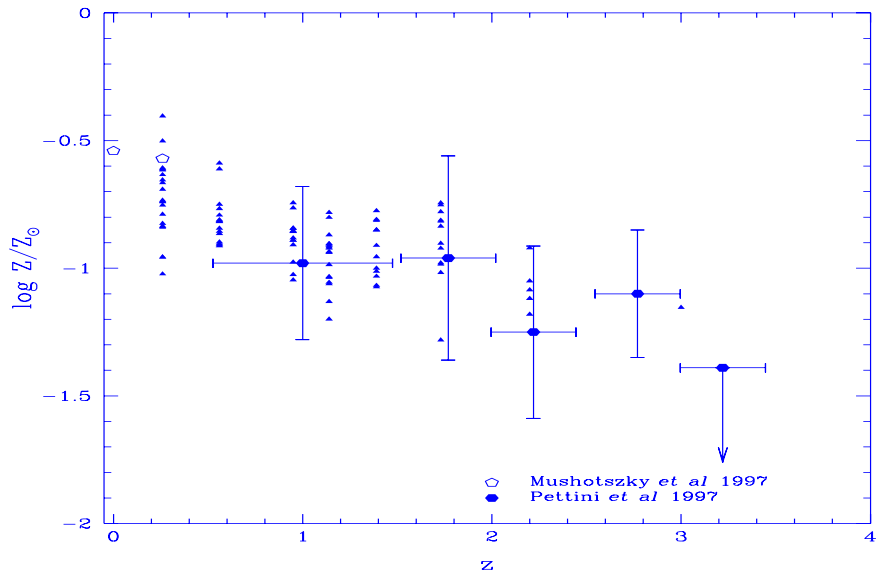}{3in}{0}{60}{40}{-190}{-74}
\caption{Metallicity of the cold gaseous component within galactic
halos as a function of redshift (filled triangles) 
compared to metal abundance in DLAs and in the intracluster medium. }
\end{figure}
For comparison we also show in this figure available data on [Zn/H] for 
DLAs systems (Pettini et al. 1997)
 and metals in the intracluster medium (Mushotszky et al. 1997).
As it can be seen, there is a very good agreement between the model
prediction and the observations, indicating that the metallicity of the
gas content of the analyzed halos of these simulations  
compare well with those of DLA systems for $z > 1$. 
We find that at lower redshifts, $z<1$, the global metallicity of the gas 
component
of these halos  show a significant increase toward
solar values, which compare with measurements of the intracluster medium.


\begin{references}
\reference 
\reference Chiappini, C., Matteucci, F. \& Gratton, R. 1997, \apj, 477,

\reference Madau, P., Pozzetti, L. \& Dickinson, M., 1998, ApJ, 498, 106

\reference Mushotszky, R. F. \& Loewenstein, M., 1997, ApJ, 481, 64L
\reference Pettini et al., 1997, ApJ, 486, 665

\reference Raiteri, C.M., Villata, M. \& Navarro, J. F., 1996, \aap, 315, 105

\reference Steinmetz, M. \& Muller, E., 1994, \aap, 281, L97

\reference Tissera, P.B., Lambas, D. G. \& Abadi, M., 1997, \mnras, 286,   




\end{references}
\end{document}